\documentclass[]{JHEP3}
\usepackage{citesort}

\newcommand{\tr}{{\mathrm{tr}}}

\newcommand{\be}{\begin{equation}}
\newcommand{\ee}{\end{equation}}
\newcommand{\bea}{\begin{eqnarray}}
\newcommand{\eea}{\end{eqnarray}}

\newcommand{\bear}{\begin{array}{l}}
\newcommand{\eear}{\end{array}}

\newcommand{\ie}{{\it i.e.}\ }

\newcommand{\eg}{{\it e.g.}\ }

\newcommand{\half}{\frac{1}{2}}

\def\eq#1{(\ref{#1})}
\def\eqs#1#2{(\ref{#1},\ref{#2})}

\def\phi{ \varphi }

\title{Equivalence of Local Potential Approximations}

\author{Tim R. Morris\\
School of Physics and Astronomy, University of
Southampton,
Highfield, Southampton SO17 1BJ, U.K.\\ E-mail:
\email{T.R.Morris@soton.ac.uk}}

\maketitle

\preprint{SHEP 0509}

\abstract{In recent papers it has been noted that the local
potential approximation of the Legendre and Wilson-Polchinski flow
equations give, within numerical error, identical results for a
range of exponents and Wilson-Fisher fixed points in three
dimensions, providing a certain ``optimised'' cutoff is used for
the Legendre flow equation. Here we point out that this is a
consequence of an exact map between the two equations, which is
nothing other than the exact reduction of the functional map that
exists between the two exact renormalization groups. We note also
that the optimised cutoff does not allow a derivative expansion
beyond second order.}

\begin{document}

The fundamentals and applications of the ``exact renormalization
group'' (exact RG), discovered and so christened independently by
Wilson and Wegner \cite{Wil,Weg}, have been studied intensively
since the beginning of the nineties \cite{Wet,Bon,erg}. The
central reason for this recrudescence is the general acceptance
that, far from being merely formal exact realizations of Wilson's
RG ideas, these ideas form the basis for powerful and flexible
approximations in non-perturbative quantum field theory. (For
reviews, see for example refs. \cite{reviews}.)

The two most widely used realizations of such exact RGs (for
others see \cite{latorre,ap}), are Polchinski's version
\cite{Pol}, equivalent to Wilson's \cite{Wil} by a change of
variables \cite{deriv,Gol}, and the version for the Legendre
effective action \cite{Wet,Bon,erg}. We will be interested in the
case where these are applied to $O(N)$ invariant $N$-component
real scalar field theory in $D$ Euclidean dimensions.

For such a theory, Polchinski's version is given
by:\footnote{We base our notation on refs. \cite{ln,erg} for
reasons that will become clear.}
\be
\label{pol}
{\partial S_\Lambda \over\partial\Lambda}={1\over2}\,
 {\delta S_\Lambda\over\delta\Phi_a}\cdot
{\partial\Delta_{UV}\over\partial\Lambda}\cdot {\delta
S_\Lambda\over\delta\Phi_a} -{1\over2}\, {\rm tr}\,
{\partial\Delta_{UV}\over\partial\Lambda}\cdot
{\delta^2S_\Lambda\over\delta\Phi\delta\Phi},
\ee
where $\Phi_a(x)$ is the $N$-component scalar field and
$S_\Lambda[\Phi]$ is the interaction part of the Wilsonian
effective action
\be
S^{eff}_\Lambda= \half\Phi_a.\Delta_{UV}^{-1}.\Phi_a +S_\Lambda.
\ee
$\Lambda$ is the effective cutoff,
$\Delta_{UV}(q,\Lambda)=C_{UV}(q,\Lambda)/q^2$ is the ultraviolet
regularised propagator, and $C_{UV}$ the ultra-violet cutoff
function.

On the other hand, the flow equation for the Legendre effective
action, also called effective average action, is given by:
\be
\label{leg}
{\partial\over\partial\Lambda}\Gamma_\Lambda[\phi] =
-{1\over2}\,\tr\ {1\over \Delta_{IR}}{\partial \Delta_{IR}\over
\partial\Lambda}\cdot A^{-1}
\ee
where
\be
A_{ab} =\delta_{ab}+\Delta_{IR}\cdot{
\delta^2\Gamma_\Lambda\over\delta\phi_a\delta\phi_b}.
\ee
Here $\Gamma_\Lambda[\phi]$ is the interaction part of the
Legendre effective action
\be
\Gamma^{tot}_\Lambda=\half\phi_a\cdot\Delta_{IR}^{-1}
\cdot\phi_a+\Gamma_\Lambda[\phi],
\ee
where the propagator has been replaced by an infrared regularised
propagator $\Delta_{IR}(q,\Lambda)= C_{IR}(q^2/\Lambda^2)/ q^2$,
$C_{IR}$ being the infrared cutoff function.

One of the simplest and most powerful approximations, which is
also widely used, is the Local Potential Approximation (LPA)
\cite{lpa}. In this case one simply makes the model approximation
that the above actions are of the form of a potential only, and
discards all parts of the right hand sides of \eq{pol} and
\eq{leg} that do not fit this approximation. More rigorously,
providing the cutoff functions are smooth, the actions have a
derivative expansion to all orders \cite{deriv}, and the LPA
simply amounts to taking the lowest order in this expansion,
setting all higher order terms to zero.

For the LPA of the Polchinski equation \eq{pol}, it turns out that
by changes of variables all explicit cutoff function dependence
disappears and thus the LPA yields universal results \cite{ball}.
For the Legendre flow equation we adopt the ``optimised" infrared
cutoff that Litim has advocated. As an additive infrared cutoff it
is \cite{litop}
\be
\label{add}
(\Lambda^2-q^2)\theta(\Lambda^2-q^2),
\ee
and thus
\bea
\Delta_{IR} &=&1/\Lambda^2\qquad {\rm for}\qquad
q<\Lambda,\nonumber\\
\label{opt}
\Delta_{IR}&=&1/q^2 \qquad\, {\rm for}\qquad q>\Lambda.
\eea
Writing $t=\ln(\mu/\Lambda)$, $\mu$ some arbitrary physical mass
scale, $S_\Lambda=\int\!d^Dx\,U(y,\Lambda)$ and
$\Gamma_\Lambda=\int\!d^Dx\,V(z,\Lambda)$, where $y=\Phi^a\Phi_a$
and $z=\phi^a\phi_a$, and scaling to dimensionless variables using
$\Lambda$ whilst also absorbing some constants,  the LPA
approximations can then be written as
\cite{ball,litop,lit}:\footnote{prime is differentiation by $y$ or
$z$ as appropriate. These are precisely the equations that appear
in ref. \cite{lit} up to scaling $z$ and $y$ by 1/2, a correction
of sign and large $N$ scaling in \eq{lpap}.}
\bea
\label{lpap}
\partial_tU(y,t)+(D-2)yU'-DU &=& - 4y(U')^2 + 2NU'
+4yU'' \phantom{{1\over V'}}\\
\label{lpal}
\partial_tV(z,t) +(D-2)zV'-DV &=& -{N-1\over1+2V'}
-{1\over1+2V'+4zV''}
\eea

We now come to the central issue of this letter. As noted first in
ref. \cite{litorig}, for the Wilson-Fisher fixed point in $D=3$
dimensions and for various $N$, several RG
eigenvalues\footnote{equivalently exponents, including those for
corrections to scaling} computed from eqs. \eqs{lpap}{lpal} agree
to all published digits \cite{litorig}. This was recently extended
by Bervillier \cite{bervil}, and analysed in much greater detail
very recently by Litim \cite{lit}. The main conclusions are that
this is a very surprising result, given the inequivalent
derivative expansions and the dependence on cutoff function
displayed by the Legendre flow equation even at the LPA level
\cite{litorig}. Most recently Litim conjectures from this ``most
remarkable [...] high degree of coincidence" that the universal
content of \eq{lpap} and \eq{lpal} must be the same \cite{lit}.
Underlining the surprising nature of the coincidence, Litim goes
on to argue that, while the derivative expansion is particularly
simple to implement in the Wilson-Polchinski equation, the
Legendre form yields more stable results even at the LPA level.

And yet, how can two partial differential equations, neither of
which is exactly soluble (even at the fixed points) yield such
non-trivial agreement, without being fundamentally related? The
answer has to be that they are in fact related by a change of
variables.

Already long ago, it was shown that as an exact statement, a
Legendre transform relation exists between the two functionals
$\Gamma_\Lambda$ and $S_\Lambda$ \cite{erg}, providing only that
the cutoff functions satisfy the sum rule
\be
\label{sum}
C_{IR}+C_{UV}=1.
\ee
The Legendre transform relation is
\be
\label{elegtr}
S_\Lambda[\Phi]=\Gamma_\Lambda[\phi]+\half
(\phi_a-\Phi_a)\cdot\Delta_{IR}^{-1}\cdot(\phi_a-\Phi_a).
\ee
It transforms the corresponding flow equations into each other
\cite{erg}: they are in fact two realizations of the same exact
RG.

Furthermore, for {\sl constant} fields $\Phi$ and $\phi$, all the
higher derivative terms inside the effective actions vanish and
this relation collapses to \cite{ln}:
\be
\label{legtr}
U(y,\Lambda)=V(z,\Lambda)+\half\Delta_{IR}^{-1}(0,\Lambda)(\phi-\Phi)^2.
\ee
Note well, that \eq{legtr} is still an exact statement. It is
trivially extended to hold for any field theory, not just $O(N)$
scalar field theory.

On the other hand, once we start to approximate the flow equations
\eqs{pol}{leg}, relation \eq{legtr} will in general be broken.
Indeed in ref. \cite{ln}, we showed that for general cutoffs,
although the large $N$ limit of the LPA for these two equations
yield the known exact results for the RG eigenvalues, only
$V(z,t)$ is exact in this limit: the Wilson-Polchinski effective
potential $U(y,t)$ is not the correct (\ie exact) potential for
general cutoffs, even in the limit of large $N$.

It is interesting to note that Litim's optimised cutoff \eq{opt}
satisfies, in particular in \eq{elegtr},
\be
\label{x}
\Delta_{IR}^{-1}=\Lambda^2\quad\hbox{\sl to all orders in the
derivative expansion.}
\ee
It is tempting to speculate that this is somehow responsible for
the preservation of the Legendre transform relation at the LPA
level. Indeed, we will shortly see that \eq{legtr} does still
hold. However, this speculation would then suggest that the
relation continues to hold at higher orders of the derivative
expansion, which seems very unlikely, and in any case \eq{opt} is
one of only many cutoff propagators with the property \eq{x},
while the functional form of \eq{lpal} is specific to the choice
\eq{opt}.

On the other hand, \eq{x} and even more importantly the lack of
smoothness in \eq{opt} (as evidenced by the appearance of
Heaviside $\theta$ function), mean the derivative expansion must
actually break down at some point. We will firm up this
observation at the end, where we will see that the derivative
expansion breaks down at $O(\partial^4)$.

One immediate corollary of these observations is that, while
\eq{lpap} is not exact for general cutoffs, even in the large $N$
limit, the large $N$ limit of this equation {\it is} the exact
answer when we use the (rather strange) cutoff following from
\eq{opt} and \eq{sum}. This thus provides a simple explanation for
why we found that \eq{lpap} nevertheless gave the correct RG
eigenvalues in the large $N$ limit.\footnote{However, we also
showed that a much larger class of incorrect equations
nevertheless give the right RG eigenvalues in this limit
\cite{ln}.}

Turning now to the proof of equivalence of the two LPAs \eq{lpap}
and \eq{lpal}, we note that the scaled version of \eq{legtr} is:
\be
\label{tr}
U(y,t)=V(z,t)+\half(\phi-\Phi)^2.
\ee
Thus
\be
\label{standard}
\partial_t|_y U = \partial_t|_z V,
\ee
a standard consequence of Legendre transforms, and equally
\be
\label{diff}
\phi_a-\Phi_a = 2\Phi_a U' = -2\phi_a V'.
\ee
From this it follows that $\phi_a$ and $\Phi_a$ point in the same
direction and thus \cite{ln}
\be
\label{sqrts}
\sqrt{z\over y} = 1-2U' = {1\over1 + 2V'}.
\ee
This relation deals with the first term on the right hand side of
\eq{lpal}. On the other hand, it also implies $U'\sqrt{y} =
V'\sqrt{z}$ and thus $zV' = yU'(1-2U')$. Squaring \eq{diff}, we
have $U-V = 2y(U')^2$, so
\be
\label{tflow}
(D-2)zV' - DV = (D-2)yU' - DU + 4y (U')^2.
\ee
Note that this concurs with the demonstration of equivalence of
\eq{pol} and \eq{leg} \cite{erg}: there, differentiating
\eq{elegtr} generates a term that cancels the tree level
contribution, \ie the first term on the right hand side of
\eq{pol}. After scaling, it is the $D$ dependent terms (generated
by the trivial $\Lambda$ dependence) that play the same r\^ole.

Differentiating \eq{sqrts} by $y$ and comparing to differentiating
the inverse by $z$, one finds
\be
1-2U'-4yU'' = {1\over1+2V'+4zV''}.
\ee
Finally, using this, \eq{standard}, \eq{sqrts} and \eq{tflow}, and
making the standard discard of the constant vacuum energy
contribution (here $-N$) \cite{Pol,erg}, it is easy to see that
\eq{lpap} and \eq{lpal} are indeed Legendre transforms of each
other under the map \eq{tr}, as claimed.

Although we established the equivalence in $O(N)$ invariant scalar
field theory, so as to make immediate contact with refs.
\cite{litorig,bervil,lit}, it is clear that the $O(N)$ invariance
is not crucial and this equivalence would follow similarly, at
least for any scalar field theory. Of course it follows that
universal information, such as the RG eigenvalues around any fixed
point, must agree between the two realizations, here \eq{lpap} and
\eq{lpal}. But more than that, the relations, \eg \eq{sqrts},
provide an explicit map between the solutions of each equation. It
is therefore only necessary to solve one of them, after which the
change of variables can be performed to get the other. Both the
solution and the change of variables can be found numerically. An
obvious but important consequence is that for the optimised cutoff
\eq{opt}, the LPA is just as accurate for all quantities
irrespective of whether we use the Wilson-Polchinski or the
Legendre form as the starting point.

We finish by completing our remark on the limitations of \eq{opt}
when considering derivative expansions. A sharp regulator is known
to break down already at $O(\partial^2)$ \cite{erg,mom}. In its
additive form \eq{add}, Litim's optimised regulator is in fact the
first integral of a sharp cutoff. We should then expect that this
regulator survives at $O(\partial^2)$, but breaks down at
$O(\partial^4)$. We now confirm this.

We can investigate the properties of momentum expansions to high
order if we first focus on perturbative contributions
\cite{tighe}. Indeed, as in that paper let us focus on the
Legendre effective action one-loop four-point vertex in $D=4$
dimensional $\lambda\phi^4$ theory. The flow of this vertex is the
sum of three contributions (the $s,t$ and $u$ channels) of the
form:
\be
-\lambda^2\int\!\!{d^4q\over(2\pi)^4}\ \Delta_{IR}(q+p,\Lambda)\,
\Lambda{\partial\over\partial\Lambda}\Delta_{IR}(q,\Lambda).
\ee
Using \eq{opt} and writing $(p+q)^2 =p^2+q^2+2pqx$, where $x$ is
the cosine of the angle between $p^\mu$ and $q^\mu$, and scaling
out $\Lambda$, the integral is proportional to
\be
\int^1_0\!\!dq\,\int^1_{-1}\!\!\!dx\ q^3 \left\{ \left(1-{1\over
(p+q)^2}\right)\theta\left[1-(p+q)^2\right] + {1\over
(p+q)^2}\right\}.
\ee
We are not interested in the proportionality constant, namely
$2\lambda^2/(4\pi)^2$. Since we will be expanding in the external
momentum $p$, we can assume it is small. The integral over $x$ is
then straightforward; the $\theta$ function is relevant only when
$q>1-p$, splitting the $q$ integral into two domains. Thus we
obtain the contribution as
\be
{1\over3}\left(p^2+{1\over
p}\right)\ln(1+p)+{1\over6}(1+p)-{4\over9}p^2-{1\over60}p^4
\ee
which expands as
\be
{1\over2}
-{1\over3}p^2+{1\over4}p^3-{7\over60}p^4+{1\over18}p^5+\cdots
\ee
If this corresponded to a derivative expansion, the powers in $p$
would all be even, however we see that as expected odd powers
appear after $O(p^2)$. Thus beyond $O(\partial^2)$ the cutoff
\eq{add}, equivalently \eq{opt}, must be treated in terms of the
more general momentum scale expansion developed in refs.
\cite{erg,mom}. However, we would also expect that, in common with
sharp cutoffs, the momentum scale expansion does not have good
convergence properties \cite{tighe}.

\end{document}